# A General Optimal Control Model of Human Movement Patterns I: Walking Gait

**Stuart Hagler**

Oregon Health & Science University

Portland, OR, USA

haglers@ohsu.edu

**Abstract:** The biomechanics of the human body gives subjects a high degree of freedom in how they can execute movement. Nevertheless, subjects exhibit regularity in their movement patterns. One way to account for this regularity is to suppose that subjects select movement trajectories that are optimal in some sense. We adopt the principle that human movements are optimal and develop a general model for human movement patters that uses variational methods in the form of optimal control theory to calculate trajectories of movement trajectories of the body. We find that in this approach a constant of the motion that arises from the model and which plays a role in the optimal control model that is analogous to the role that the mechanical energy plays in classical physics. We illustrate how this approach works in practice by using it to develop a model of walking gait, making all the derivations and calculations in detail. We finally show that this optimal control model of walking gait recovers in an appropriate limit an existing model of walking gait which has been shown to provide good estimates of many observed characteristics of walking gait.

## 1 Introduction

The biomechanics of the human body give a subject a high degree of freedom in how a movement may be executed. Nevertheless, subjects exhibit great regularity in their movement patterns. One way to account for this observed regularity is to suppose that subjects regularly choose movements that are optimal in some rigorous sense. Variational methods can generate movement trajectories by finding the trajectory that is optimal in the sense of minimizing some measure of cost. These methods are a standard part of the toolkit physicists use calculate physical trajectories in classical mechanics. In this paper, we adopt a variational approach to creating models of human movement that generate movement trajectories body. We do this by generalizing the kinds of variational methods used by physicists by using the mathematics of optimal control theory to calculate trajectories for movements of the body. The aim of the optimal control approach to modeling human movements is to produce a model that can generate trajectories for arbitrary movements of the human body. We suppose that all human movements are variations on a common motor control process only differing in the part of the body being used and the goals being met. We therefore believe that we should study human movements beginning with a modeling approach which can be used to construct the trajectories for any human movement and then look at how the model may need to be modified to correctly describe specific movements. However, as it makes things clearer to illustrate the optimal control modeling approach using a specific movement, we illustrate the approach in this paper using walking gait.

We can think of the problem of the optimal control of the human body in the following terms. The human body is a skeleton that is moved by the actions of muscles generating forces at points on the skeleton.

A movement takes the skeleton from one pose into another over some interval of time. We can identify a small number of points on the body as being the points whose trajectories are controlled during the movement and constrain the rest of the body so that it must move in some way that is consistent with the motion of the controlled points as in [1-5]. The points themselves are controlled by setting the n-th time derivative of their position to a sequence of arbitrary values throughout the time interval in which the movement occurs; these n-th time derivatives of the positions are the *controls* of the body. We choose the sequences of values for the controls by finding those that generate a movement of the skeleton that minimizes some measure of cost associated with the movement.

One choice of control is the *jerk* (the third time-derivative of the position). It has been invoked as a general principle for understanding human movements and has been used as the basis for calculating movement trajectories. [6-11] We have previously used optimal control models built on jerk-control to describe vertical movements of the torso in the course of a simple squatting exercise [12] and movements of the hand to control a computer mouse. [13] In constructing the optimal jerk-control model of the computer mouse in [13] we developed the mathematical apparatus of optimal jerk-control models in detail. The choice of the jerk as the control means that while the jerk can be set to an arbitrary sequence of values over the interval in which the movement occurs, the position, velocity, and acceleration are continuous functions of time. Looking to Newton's second law, we see that having the acceleration be continuous in time means that the forces acting on the controlled points of the body must be continuous in time. Therefore, choosing jerk as the control implicitly means constructing a model in which the forces generated by the muscles are continuous in time. Again, looking to Newton's second law, we see that controlling the jerk corresponds to controlling the *yank* (the first time-derivative of the force) of the muscles. Thus, a model in which the jerk is controlled looks like a more conventional optimal control model in which the yank is controlled when the model is reformulated in terms of the muscle forces.

The key component of an optimal control model is the cost functional which assigns a cost to every possible movement trajectory. The cost functional contains the body of physical theory that describes the system while finding the solution that minimizes the cost functional is simply a matter of established mathematics. We must therefore be clear about how the cost functional is constructed in the model we develop. Our approach is to begin with what we take to be the most unobjectionable supposition about human movement – namely, that, all other things being equal, subjects will move in a way that minimizes the metabolic energy expended in executing the movement. In [14-16], we constructed and validated a model that estimates the metabolic energy expended in executing a walking gait. This model estimates the metabolic energy using the values of the muscle forces required to execute the walking gait. However, the model in [14-16] only uses the forces and not the yanks, and we have argued that the optimal control model should be formulated in terms of control of the yanks. We therefore associate a cost with the yanks of a movement and introduce terms in the yank that are mathematically convenient for the calculation of the movement trajectories.

In this paper, we illustrate the general optimal control approach to modeling human movement patterns using walking gait. In [14-16], we constructed a simple biomechanical model of walking gait using the principle of movement optimality that could account for several characteristics of walking gait, namely: (i) the metabolic energy expenditure per step for the range of avg. walking speeds and avg. step lengths observed by Atzler & Herbst, [17] (ii) the regular pattern of combinations of avg. walking speed and avg. step length selected by subjects as observed by Grieve, [18] (iii) the lower limit of avg. walking speeds for very slow walking gaits selected by pre-rehabilitation Parkinson's disease subjects observed by Frazzitta et al. [19] and for very slow walking gaits by older adults by Hagler et al., [20] and (iv) the functional forms of measures of the energy efficiency across a range of walking gaits observed by Donovan & Brooks. [21] Since this model moves the body using fixed trajectories, it only describes walking gait down to scales of about a step. As the optimal control model of walking gait that we construct in this paper generates movement trajectories, we expect that it provides a means of extending the model in [14-16] to a model able to describe walking gait at scales below that of a step. We therefore take an interest in wedding the model in [14-16] to that constructed here both to show that we can extend that model and to show that we can recover established results for walking gait from [14-16] using this model. We do this by showing that we can recover the model in [14-16] as a limiting case of the optimal control model of walking gait.

This paper is structured as follows. We first provide some definitions and summarize the modeling of movement and metabolic energy we developed for walking gait in [14-16], show how we, in general, propose to construct cost functionals to describe human movements using optimal control theory, and outline the mathematics used to derive movement trajectories from these cost functionals (Sec. 2). We construct the optimal control model for walking gait by generalizing the model that we used in [14-16] from specified to arbitrary trajectories of movement for skeleton, plugging the resulting model into a cost functional, and finding the optimal trajectories (Sec. 3). We finally show that the trajectories of the torso and swing foot in the optimal control model can be made to approximate those used in the model in [14-16] in the case of symmetric and uniform walking gait in the limit of low yanks (Sec. 4). We therefore find that we can generalize the model developed in [14-16] to scales below that of a step using an optimal control model.

## 2 Optimal Control Model

We begin by outlining the optimal control approach to modeling human movements that we adopt in this paper. We first define some convenient terminology and mathematical notation. We then review the model that we developed in [14] to describe movements of the body; it is composed of three parts: (i) the segment model that gives the skeleton that we use to describe the poses of the body, (ii) the model we use to describe the metabolic energy of a movement, and (iii) the model we use to describe the perceived muscle forces. Working from the relationship between the muscle forces and the metabolic energy of the movement, we develop the cost functional that we use to generate movement trajectories. We finally provide an outline of the mathematics that we use to calculate the optimal trajectories given a cost functional for a movement.

*2.1 Terminology & Notation*

For a position variable $x(t)$, the first- and second-order time-derivatives are the velocity $v(t) = \dot{x}(t)$ and the acceleration $a(t) = \ddot{x}(t)$. We follow [13] and denote the higher order time-derivatives as the *jerk* $\dddot{x}(t)$, the *snap* $\ddddot{x}(t)$, the *crackle* $\dddddot{x}(t)$, and the *pop* $\ddddddot{x}(t)$. For a force variable $F(t)$ (e.g. example representing the net force of the muscles acting on a point on the body as a function of time), we denote the first-order time-derivative as the *yank* $\dot{F}(t)$. We denote an ordinary function as $\xi(\cdot)$ using parentheses, and a functional (a function of a function) as $\xi[\cdot]$ using brackets. We denote the set of all values $\xi_n$ as $\{\xi_n\}$. For a movement taking a time $T$, and beginning at time $-T/2$ and ending at time $T/2$, the we denote the time average of a variable $\xi$ associated with the movement by $\langle \xi \rangle = (1/T) \int_{-T/2}^{T/2} \xi dt$. A function $\xi(t)$ is *odd* if $\xi(t) = -\xi(-t)$, and it is *even* if $\xi(t) = \xi(-t)$. We find it contributes to the clarity to distinguish the metabolic and mechanical energies of a movement of how the metabolic energy (i.e. the stored energy the body consumes when activating muscles) and mechanical energy (i.e. the kinetic and potential energies that describe the trajectories of the segments of the body as they move through space) by expressing them in different units.

We express the metabolic energy using calories and mechanical energy using joules; these two measures of energy are related to each other by 1.0 cal = 4.2 J.

*2.2 Segment, Metabolic Energy, & Perceived Muscle Force Models*

We model the skeleton of the human body as in [14-16] using a segment model consisting of a system of N ("nu") segments attached at $N$ joints. An example of such a model can be found in [22] using segments for the feet, lower legs, upper legs, hands, lower arms, upper arms, torso, and head. A movement is described by a set of trajectories $\vec{x}_n(t)$ of points on the body and the rest of the body is constrained to move sensibly given the motion of this specified set of points. Associated with each controlled point on the body is an effective mass $m_n$, a trajectory $\vec{x}_n(t)$, and a net muscle force $\vec{F}_n(t)$ that determines the acceleration of that point according to Newton's second law using the effective mass.

The metabolic rate $\dot{W}(t)$ of expending metabolic energy is the sum of metabolic rates $\dot{W}_n(t)$ associated with each joint $\dot{W}(t) \approx \sum_{n=1}^{N} \dot{W}_n(t)$. Following [14], the metabolic rate associated with a joint is given by a function:

$$\dot{W}_n(t) = \dot{W}_n^F\left(\vec{F}_n(t)\right) + \dot{W}_n^E\left(\vec{F}_n(t), \vec{v}_n(t)\right). \tag{1}$$

Further continuing the argument in [14], the metabolic rates $\dot{W}_n^F(t)$ and $\dot{W}_n^E(t)$ take on approximate mathematical forms given by the lowest order of a Taylor series expansion given some physical constraints that we do not give here; the mathematical forms are:

$$\begin{aligned}\dot{W}_n^F\left(\vec{F}_n(t)\right) &\approx \varepsilon_n F_n(t)^2, \\ \dot{W}_n^E\left(\vec{F}_n(t), \vec{v}_n(t)\right) &\approx \eta_n \vec{F}_n(t) \cdot \vec{v}_n(t).\end{aligned} \tag{2}$$

The metabolic rates $\dot{W}(t)$, $\dot{W}_n(t)$, $\dot{W}_n^F(t)$, and $\dot{W}_n^E(t)$ have dimensions of metabolic power $[cal \cdot s^{-1}]$. The quantities $\varepsilon_n$ and $\eta_n$ are constant parameter values characterizing the associated metabolic rates. The parameters $\eta_n$ may take on different values when the muscles add or remove mechanical energy to or from a segment, though we require they be constant in each case. The metabolic energies $W_n^F[\vec{F}_n]$ associated with generating the muscle forces in a segment are:

$$W_n^F\left[\vec{F}_n\right] = \int_{-T/2}^{T/2} \varepsilon_n F_n^2 dt = \varepsilon_n \left\langle F_n^2 \right\rangle T. \tag{3}$$

The metabolic energies $W_n^F$ have dimensions of metabolic energy $[cal]$.

In [14], we gave a simple model for the perceived muscle force $\psi_{F,n}[\vec{F}_n]$ for the subject over the course of a movement in which it is approximately related to the muscle force $\vec{F}_n(t)$ as:

$$\psi_{F,n}\left[\vec{F}_n\right] \approx \left(1 / F_{0,n}^2\right)\left\langle F_n^2 \right\rangle. \tag{4}$$

Here $F_{0,n}$ is a constant with dimensions of force $[N]$ so $\psi_{F,n}$ is dimensionless. In the static case where there is a constant muscle force $\vec{F}_n$ supporting a weight without moving the body, the perceived muscle force $\psi_{F,n}[\vec{F}_n]$ approximates to Stevens' power law for this case ($\psi_{F,n} \approx \left(|\vec{F}_n|/F_{0,n}\right)^{1.7}$). [23, 24]

*2.3 Cost Functional*

The most unobjectionable choice for the cost functional $J[\cdot]$ of a movement is the metabolic energy expended in executing the movement. In this case, the model would select the movement trajectory that minimizes the metabolic energy. However, if we assume that subjects possess a mechanism by which they can perceive the cost of a movement, then we find that we cannot use the total metabolic energy expended. The model that we have given in Sec. 2.2 provides a mechanism by which subjects can perceive the magnitudes of forces but does not provide a similar mechanism to perceive the total metabolic energy. Instead, we can use the model for the perception of muscle forces in (4) to construct an estimator for the portion of the metabolic energy expended generating the muscle forces. This would allow us to construct the cost functional would be $J[\{\vec{F}_n\}] = W^F[\{\vec{F}_n\}] \approx \left(\sum_n(\varepsilon_n F_{0,n}^2)\psi_{F,n}[\vec{F}_n]\right)T$. However, we have argued that the optimal control model should involve control of the yanks of the muscles and the yank appears nowhere in this cost functional. We must include terms in the yank in the cost functional.

We need to construct terms of the form $j_n(\vec{\dot{F}}_n)$ associating a cost with a yank $\vec{\dot{F}}_n$. We assume that the cost is zero when the yank is zero, and the cost is independent of the direction of the yank so that the term takes the form $j_n(\dot{F}_n^2)$. Taking a Taylor series expansion of this and truncating to the lowest order, we find $j_n(\dot{F}_n^2) \approx \alpha_n \dot{F}_n^2$ for some constant parameter value $\alpha_n$. Thus, the cost associated with the yanks for a movement takes the form $\int_{-T/2}^{T/2}\sum_n j_n(\dot{F}_n^2)\,dt \approx \int_{-T/2}^{T/2}\sum_n \alpha_n \dot{F}_n^2 dt$. To arrive at a cost functional in which the jerk is controlled for a movement taking a beginning at time $-T/2$ and ending at time $T/2$, use a cost functional $J\left[\{\vec{F}_n, \vec{\dot{F}}_n\}\right]$ given by:

$$J\left[\left\{\vec{F}_n, \vec{\dot{F}}_n\right\}\right] = \left(\sum_n \left(\varepsilon_n F_{0,n}^2\right)\psi_{F,n}\left[\vec{F}_n\right]\right)T + \int_{-T/2}^{T/2}\sum_n \alpha_n \dot{F}_n^2 dt. \quad (5)$$

We note that the cost $J$ has dimensions of metabolic energy $[cal]$.

We can rewrite the yank-control cost functional $J\left[\{\vec{F}_n, \vec{\dot{F}}_n\}\right]$ in (5) as:

$$J\left[\left\{\vec{F}_n, \vec{\dot{F}}_n\right\}\right] = \int_{-T/2}^{T/2}\sum_n \left(\alpha_n \dot{F}_n^2 + \varepsilon_n F_n^2\right)dt. \quad (6)$$

As we want to use (6) to derive the trajectory that minimizes the cost, we must further rewrite (6) in terms of the trajectory. We expect that each of the forces $\vec{F}_n(t)$ can be expressed as a function of the set of functions $\{\vec{x}_n, \vec{\dot{x}}_n, \vec{\ddot{x}}_n, \vec{\dddot{x}}_n\}$ associated with the trajectories of the various points on the body that the subject controls and that we can rewrite the yank-control cost functional $J\left[\{\vec{F}_n, \vec{\dot{F}}_n\}\right]$ in terms of the set of functions $\{\vec{x}_n, \vec{\dot{x}}_n, \vec{\ddot{x}}_n, \vec{\dddot{x}}_n\}$ as the jerk-control cost functional:

$$J\left[\left\{\vec{x}_n, \vec{\dot{x}}_n, \vec{\ddot{x}}_n, \vec{\dddot{x}}_n\right\}\right] = \int_{-T/2}^{T/2} L\left(\left\{\vec{x}_n, \vec{\dot{x}}_n, \vec{\ddot{x}}_n, \vec{\dddot{x}}_n\right\}\right)dt. \quad (7)$$

*2.4 Lagrangian Method of Solving the Optimal Control Problem*

One way to calculate the trajectory that minimizes the cost functional in (7) using the Lagrangian method of solving the optimal control problem. This method begins by identifying the *Lagrangian* of the system; it is the integrand $L(\{\vec{x}_n, \vec{\dot{x}}_n, \vec{\ddot{x}}_n, \vec{\dddot{x}}_n\})$ in (7) and has dimensions of metabolic rate $[cal \cdot s^{-1}]$. We must first identify all the independent components that make up the Lagrangian by specifying the scalar components that make up the vector $\vec{x}_n(t)$ as $x_{m,n}(t)$. Once we have done this, as we have shown in [13], we can then find the trajectories $x_{m,n}(t)$ that optimize the value of the cost functional in (7) by solving the system of differential equations given by:

$$\left\{ \frac{\partial L}{\partial x_{m,n}} - \frac{d}{dt}\frac{\partial L}{\partial \dot{x}_{m,n}} + \frac{d^2}{dt^2}\frac{\partial L}{\partial \ddot{x}_{m,n}} - \frac{d^3}{dt^3}\frac{\partial L}{\partial \dddot{x}_{m,n}} = 0 \right\}. \tag{8}$$

Equation (8) gives a critical point for the system and it remains a further step to prove that the critical point is a minimum. For most cost functionals that we would apply to human movements we expect a critical point would also be a minimum.

*2.5 Hamiltonian Methods of Solving the Optimal Control Problem*

A second way to the trajectory that minimizes the value of the cost functional in (7) is by using the Hamiltonian method of solving the optimal control problem. This method begins by defining the *generalized coordinate vectors $Q_{m,n}$, controls $u_{m,n}$,* and *generalized momentum vectors $P_{m,n}$* as:

$$\begin{aligned} Q_{m,n}^{\mathrm{T}} &= \begin{bmatrix} x_{m,n} & \dot{x}_{m,n} & \ddot{x}_{m,n} \end{bmatrix}, \\ u_{m,n} &= \dddot{x}_{m,n}, \\ P_{m,n}^{\mathrm{T}} &= \begin{bmatrix} p_{1,m,n} & p_{2,m,n} & p_{3,m,n} \end{bmatrix}. \end{aligned} \tag{9}$$

We note that the generalized coordinate vectors $Q_{m,n}$ has been defined so that the various elements have different dimensions; it can be made to have the same dimensions for all its elements through the introduction of appropriately dimensioned constants. We have discussed the interpretation of the elements of the generalized momentum vectors $P_{m,n}$ in [13]. The *Hamiltonian H* is obtained by taking the Legendre transform of the Lagrangian. In our case, this amounts to combining the Lagrangian in (7) and the generalized coordinate vectors $Q_{m,n}$ and generalized momentum vectors $P_{m,n}$ in (9) as shown in [13], namely:

$$H = \sum_{m,n} P_{m,n}^{\mathrm{T}} \dot{Q}_{m,n} - L. \tag{10}$$

The Hamiltonian $H$ has dimensions of metabolic rate $[cal \cdot s^{-1}]$. Since we must require the $P_{m,n}^{T} \dot{Q}_{m,n}$ to have dimensions the same dimensions as the Lagrangian and the Hamiltonian, we require the components of the generalized momentum vector $P_{m,n}$ to have the necessary dimensions to make that true. As we have shown in [13], we can find the trajectories $x_{m,n}(t)$ that minimize the value of the cost functional in (7) by solving the system of differential equations given by:

$$\begin{aligned} &\left\{ \dot{Q}_{m,n} = \partial H / \partial P_{m,n} \right\}, &(A) \\ &\left\{ \dot{P}_{m,n} = -\partial H / \partial Q_{m,n} \right\}, &(B) \\ &\left\{ \partial H / \partial u_{m,n} = 0 \right\}. &(C) \end{aligned} \tag{11}$$

Pontryagin's minimum principle (see e.g., [25] Chapter 5) tells us that the Hamiltonian takes a constant value when calculated along the optimal trajectory (i.e. the trajectory that solves the systems in (8) and (11)). We call this constant value the *generalized energy* $\Psi$, so along the optimal trajectory we find:

$$H(t) = \Psi. \tag{12}$$

The generalized energy $\Psi$ has dimensions of metabolic rate $[cal \cdot s^{-1}]$. The generalized energy $\Psi$ plays a role in our optimal control model which is analogous to the role the mechanical energy plays in classical mechanics due to the fact that it is conserved over the course of an optimal movement as indicated in (12) . Along an optimal trajectory it gives a constant of the motion that provides information about the entire trajectory.

## 3 Optimal Control Model of Walking Gait

We now use the general optimal control approach to modeling human movements outlined in Sec. 2 to construct a model of walking gait. We first provide some useful anthropometric values. We then proceed using the same segment model for walking gait that we used in [14-16]; namely a simple two segment model consisting of two legs attached at single point representing the "torso." However, where we used specified trajectories of the body in [14-16], in the present treatment, we generalize the model to use arbitrary trajectories. We construct a yank-control cost functional in terms of these muscle force trajectories, which we then rewrite as a jerk-control cost functional in terms of the body segment trajectories. We then find the optimal trajectories of the body using the Lagrangian method of the optimal control problem and calculate the generalized energies for the trajectories using the Hamiltonian method. We conclude by looking at the mechanical energy of walking gait.

*3.1 Some Anthropometric Values*

A subject with mass $M$ and height $H$ has a mass in each leg (i.e. thigh, shank, and foot) of about $\mu = rM$, and the length of the leg of about $L = \rho H$ where $r = 0.16$ and $\rho = 0.53$. [22] The mass of the torso carried by the stance leg during a step is $m = (1 - 2r)M$.

*3.2 Walking Gait Model*

We model the body using a two-segment model with one segment for each leg — the legs are straight, do not bend at the knee, but can change length. The mass of the torso is placed in the *torso* which is the point where the two segments meet; the mass of each leg is placed in the foot at the far end of the leg segments from the torso. During a step, the torso maintains a constant height while the swing foot rises only a negligible height above the ground. We include a discontinuity at heel-strike. During one step, one leg is the *stance leg* which supports the torso as the torso moves over it, and the other is the *swing leg* which swings under the torso; the feet of the two legs are the *stance foot* and *swing foot*, respectively. The stance foot remains fixed on the ground while the swing moves to the next position; the legs lengthen or shorten as needed by the movement — we intend the amount of lengthening and shortening to be consistent with reasonable knee-bending during walking. We define the unit vector $\hat{v}$ to be the direction of motion of the torso. We describe steady state walking gaits already in progress (i.e. ignoring starting and stopping) using two gait parameters: (i) the avg. walking speed $v$ of the torso, and (ii) the avg. step length $s$ that corresponds to the distance the torso travels each step. The gait parameter $v$ giving the walking speed should not be confused with the velocities $\vec{v}_n$ appearing in the metabolic energy model or the unit vector $\hat{v}$; the diacritical mark or the absence of one suffices to distinguish them.

We assume that the metabolic energy associated with holding the body up against gravity is approximately constant over all reasonable trajectories of body for a walking gait and does not contribute to the cost of a walking gait. We allow that some amount of mechanical energy may be conserved during walking although we do not specify the mechanism. We allow the subject to apply an external force $f_{ext}(t)$ to an external object – in the case analyzed in this paper, by pulling a cart. We may treat the external force as an independent variable or constraining it in some way to be a function of the motion of the torso as $f_{ext}(\{x_{torso}, \dot{x}_{torso}, \ddot{x}_{torso}, \dddot{x}_{torso}\})$. For example, in the case of walking while pulling a cart, we may want to constrain the cart to maintain a constant position relative to the torso. In practice, since the cart is pulled by the arm, its position relative to the torso may vary somewhat due to motion of the arm. We choose, therefore, to solve for the case where the external force $f_{ext}(t)$ is not a function of the trajectory and allow the position of the cart to vary relative to the torso. We denote the muscle force applied by the stance leg to the torso by $\vec{F}_{st}(t)$ and the force applied by the swing leg to the swing foot by $\vec{F}_{sw}(t)$; we assume the stance foot is fixed on the ground during the step. When the subject applies an external force $f_{ext}(t)$ to pull some object, there is a force opposing the horizontal motion of the torso so that it can be written $-f_{ext}(t)\hat{v}$. The force the body must apply to compensate for the external force is $f_{ext}(t)\hat{v}$.

We look first at the muscle forces $\vec{F}_{st}(t)$ and $f_{ext}(t)$ applied by the stance leg to the torso as illustrated in Fig 1. During the first half of the step, gravity pulls the torso in the direction $-\hat{v}$, while, during the second half, gravity pulls the torso in the direction $\hat{v}$. The torso already has the mechanical energy needed to carry it forward with the required speed, so the muscle forces counteract the force of gravity and do external mechanical work.

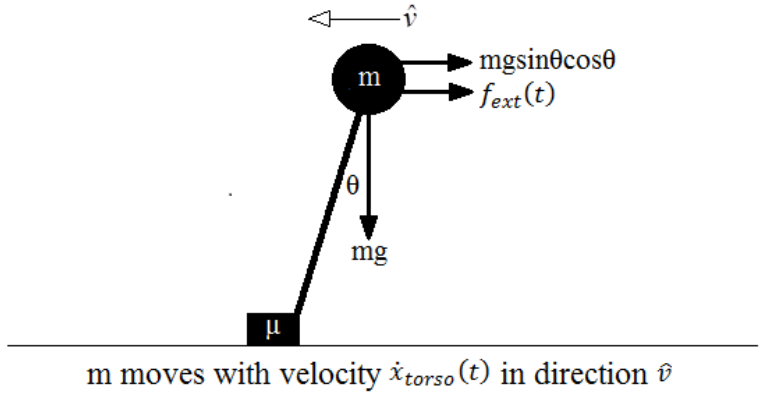

**Figure 1.** The motion of the torso over the stance leg during one step of walking gait. The torso moves over the leg at a speed $\dot{x}_{torso}$ while producing and external force $f_{ext}$. The muscles must provide the necessary force to compensate for the effect of gravity on the torso's speed.

The angle $\theta(t)$ of the stance leg with respect to the vertical determines the effect of gravity on the torso. We define $\theta(t)$ so that it is negative during the first half of the step, and positive during the second half. The horizontal force related to the torque on the inverted pendulum by gravity is $mg\sin\theta\cos\theta = (1/2)mg\sin2\theta$ where $g$ denotes the acceleration of gravity; we find that $\theta(t)$ and $\vec{F}_{st}(t)$ satisfy:

$$\begin{aligned} L\sin\theta &= x_{torso}, \\ \vec{F}_{st} &= \left(m\ddot{x}_{torso} - (1/2)mg\sin2\theta + f_{ext}\right)\hat{v}. \end{aligned} \quad (13)$$

For small angles $\theta$ corresponding to small avg. step lengths $s$, we can take $(1/2)\sin2\theta \approx \sin\theta$:

$$\begin{aligned} L\sin\theta &= x_{torso}, \\ \vec{F}_{st} &\approx \left(m\ddot{x}_{torso} - mg\sin\theta + f_{ext}\right)\hat{v}. \end{aligned} \quad (14)$$

Thus, in the case of small avg. step lengths $s$, the force $\vec{F}_{st}(t)$ of the stance leg on the torso is approximately given by:

$$\vec{F}_{st} \approx \left(m\ddot{x}_{torso} - (mg/L)x_{torso} + f_{ext}\right)\hat{v}. \quad (15)$$

We look next at the muscle forces $\vec{F}_{sw}(t)$ applied by the torso to the swing leg as illustrated in Fig. 2. We require the swing foot to move horizontally with acceleration $\ddot{x}_{foot}(t)$. During one step, the body moves one avg. step length, and the swing foot travels one stride length horizontally in direction $\hat{v}$. The swing leg begins and ends the swing at rest, and so must accelerate and decelerate as required over the course of the swing. During the first half of the step, gravity pulls the swing leg in the direction $\hat{v}$, while, during the second half, gravity pulls the swing leg in the direction $-\hat{v}$.

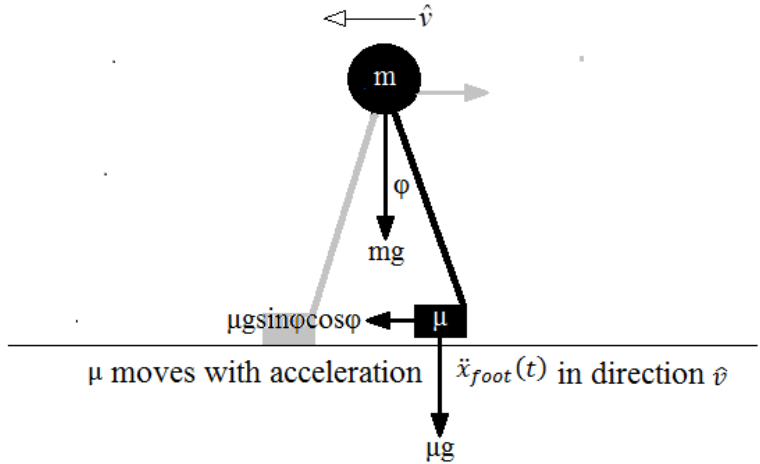

**Figure 2.** The motion of the swing leg under the torso during one step of walking gait. The swing leg moves symmetrically under the torso, accelerating with an acceleration $\ddot{x}_{foot}$. The muscles must provide the force that generates the acceleration and compensate for the effect of gravity on the swing leg.

The angle $\varphi(t)$ of the swing leg with respect to the vertical determines the effect of gravity on the swing leg. We define $\varphi(t)$ so that it is negative during the first half of the step, and positive during the second half. As the acceleration of the swing leg is $\ddot{x}_{foot}(t)$. The horizontal muscle forces that must be applied to the leg to generate the acceleration is $\mu \ddot{x}_{foot} + \mu g \sin\varphi\cos\varphi$; we find that $\varphi(t)$ and $\vec{F}_{sw}(t)$ for the first half of the step satisfy:

$$\begin{aligned} L\sin\varphi &= x_{foot} - x_{torso}, \quad -T/2 \le t \le T/2, \\ \vec{F}_{sw} &= \left(\mu \ddot{x}_{foot} + (\mu g/2)\sin 2\varphi\right)\hat{v}. \end{aligned} \tag{16}$$

For small angles $\theta$ corresponding to small avg. step lengths $s$, we can take $(1/2)\sin 2\theta \approx \sin\theta$:

$$\begin{aligned} L\sin\varphi &\approx x_{foot} - x_{torso}, \\ \vec{F}_{sw} &\approx \left(\mu \ddot{x}_{foot} + \mu g \sin\varphi\right)\hat{v}. \end{aligned} \tag{17}$$

Thus, in the case of small avg. step lengths $s$, the force $\vec{F}_{sw}(t)$ of the swing leg on the swing foot is approximately given by:

$$\vec{F}_{sw} \approx \left(\mu \ddot{x}_{foot} + (\mu g/L)(x_{foot} - x_{torso})\right)\hat{v}. \tag{18}$$

*3.3 Cost Functional*

The cost functional for the two-segment model of walking gait is (cf. (6)):

$$J = \int_{-T/2}^{T/2} \left(\alpha_{st}\dot{F}_{st}^2 + \alpha_{sw}\dot{F}_{sw}^2 + \varepsilon_{st}F_{st}^2 + \varepsilon_{sw}F_{sw}^2\right)dt. \tag{19}$$

To keep the expression of the cost functional and the resulting expressions derived from it relatively uncluttered, we define three frequency parameter values $\omega_{st}$, $\omega_{sw}$, and $\omega_{sp}$ given by:

$$\begin{aligned} \omega_{st}^2 &= \varepsilon_{st}/\alpha_{st}, \\ \omega_{sw}^2 &= \varepsilon_{sw}/\alpha_{sw}, \\ \omega_{sp}^2 &= g/L. \end{aligned} \tag{20}$$

Using these frequency parameters, the cost functional $J$ may be rewritten in terms of the trajectory $x_{torso}(t)$ of the torso, the trajectory $x_{foot}(t)$ of the swing foot, and the trajectory $f_{ext}(t)$ of the external force as:

$$\begin{aligned} J = \ &\alpha_{st}m^2 \int_{-T/2}^{T/2} \left(\dddot{x}_{torso} - \omega_{sp}^2 \dot{x}_{torso} + \dot{f}_{ext}/m\right)^2 dt \\ &+\alpha_{sw}\mu^2 \int_{-T/2}^{T/2} \left(\dddot{x}_{foot} + \omega_{sp}^2(\dot{x}_{foot} - \dot{x}_{torso})\right)^2 dt \\ &+\varepsilon_{st}m^2 \int_{-T/2}^{T/2} \left(\ddot{x}_{torso} - \omega_{sp}^2 x_{torso} + f_{ext}/m\right)^2 dt \\ &+\varepsilon_{sw}\mu^2 \int_{-T/2}^{T/2} \left(\ddot{x}_{foot} + \omega_{sp}^2(x_{foot} - x_{torso})\right)^2 dt. \end{aligned} \tag{21}$$

To calculate the general optimal trajectory for walking gait using the cost functional in (21) would require that we solve a single system of differential equations combining the trajectories of $x_{torso}(t)$, $x_{foot}(t)$, and $f_{ext}(t)$. If we restrict our attention to normal, healthy walking gaits, we can simplify the problem of finding the optimal trajectory somewhat. We expect that for normal, healthy walking gaits the torso will move in a way that is reasonably approximated by a constant horizontal velocity $x_{torso}(t) \approx vt$. We can rewrite this as:

$$x_{foot} - x_{torso} \approx x_{foot} - vt. \tag{22}$$

Substituting (22) into (21), we obtain an approximate cost functional that describes normal, healthy walking gaits:

$$\begin{aligned}
J \approx{}& \alpha_{st} m^2 \int_{-T/2}^{T/2} \left( \dddot{x}_{torso} - \omega_{sp}^2 \dot{x}_{torso} + \dot{f}_{ext} / m \right)^2 dt \\
& + \alpha_{sw} \mu^2 \int_{-T/2}^{T/2} \left( \dddot{x}_{foot} + \omega_{sp}^2 \left( \dot{x}_{foot} - v \right) \right)^2 dt \\
& + \varepsilon_{st} m^2 \int_{-T/2}^{T/2} \left( \ddot{x}_{torso} - \omega_{sp}^2 x_{torso} + f_{ext} / m \right)^2 dt \\
& + \varepsilon_{sw} \mu^2 \int_{-T/2}^{T/2} \left( \ddot{x}_{foot} + \omega_{sp}^2 \left( x_{foot} - vt \right) \right)^2 dt.
\end{aligned} \tag{23}$$

The approximate cost functional in (23) consists of a linear combination of terms either involving the trajectory of the torso and the external force, or the trajectory of the swing foot. We may therefore separate the cost functional in (23) into two cost functionals that can be solved independently. The first cost functional is $J_{st}$ measures the cost of moving the torso over the stance leg while applying the external force; it is:

$$\begin{aligned}
J_{st} ={}& \alpha_{st} m^2 \int_{-T/2}^{T/2} \left( \dddot{x}_{torso} - \omega_{sp}^2 \dot{x}_{torso} + \dot{f}_{ext} / m \right)^2 dt \\
& + \varepsilon_{st} m^2 \int_{-T/2}^{T/2} \left( \ddot{x}_{torso} - \omega_{sp}^2 x_{torso} + f_{ext} / m \right)^2 dt.
\end{aligned} \tag{24}$$

The second cost functional is $J_{sw}$ measures the cost of moving the swing foot under the torso; it is:

$$\begin{aligned}
J_{sw} ={}& \alpha_{sw} \mu^2 \int_{-T/2}^{T/2} \left( \dddot{x}_{foot} + \omega_{sp}^2 \left( \dot{x}_{foot} - v \right) \right)^2 dt \\
& + \varepsilon_{sw} \mu^2 \int_{-T/2}^{T/2} \left( \ddot{x}_{foot} + \omega_{sp}^2 \left( x_{foot} - vt \right) \right)^2 dt.
\end{aligned} \tag{25}$$

*3.4 Optimal Trajectories of the Torso & External Force*

The optimal trajectories of the torso and external force minimize the cost functional $J_{st}$ in (24). We first calculate the optimal trajectories using the Lagrangian method of solving the optimal control problem. This gives a system of differential equations that separates into two problems: (i) calculating the optimal trajectory of the torso over the stance leg, and (ii) calculating the optimal trajectory of the external force. We then calculate constants of the motion in the form of the generalized energies using the Hamiltonian method of solving the optimal control problem.

*3.4.1 Lagrangian Method*

The Lagrangian $L_{st}$ governing the motion of the torso and external force is the integrand of the cost functional $J_{st}$ in (24); that is:

$$L_{st} = \alpha_{st} m^2 \left( \dddot{x}_{torso} - \omega_{sp}^2 \dot{x}_{torso} + \dot{f}_{ext}/m \right)^2 \\ + \varepsilon_{st} m^2 \left( \ddot{x}_{torso} - \omega_{sp}^2 x_{torso} + f_{ext}/m \right)^2. \tag{26}$$

Since we are treating the external force $f_{ext}(t)$ independently of the trajectory $x_{torso}(t)$ of the torso, a system of two differential equations is obtained from the Lagrangian $L_{st}$ using (8). The first differential equation of the system is obtained taking the partial derivatives of $L_{st}$ using the various time-derivatives of $x_{torso}(t)$; it is:

$$\left( \left( \dddddot{x}_{torso} - \omega_{sp}^2 \ddddot{x}_{torso} \right) - \omega_{st}^2 \left( \dddot{x}_{torso} - \omega_{sp}^2 \ddot{x}_{torso} \right) \right) \\ -\omega_{sp}^2 \left( \left( \dddot{x}_{torso} - \omega_{sp}^2 \ddot{x}_{torso} \right) - \omega_{st}^2 \left( \ddot{x}_{torso} - \omega_{sp}^2 x_{torso} \right) \right) \\ + m^{-1} \left( \dddot{f}_{ext} - \omega_{st}^2 \ddot{f}_{ext} \right) - \omega_{sp}^2 m^{-1} \left( \ddot{f}_{ext} - \omega_{st}^2 f_{ext} \right) = 0. \tag{27}$$

The second differential equation of the system is obtained taking the partial derivatives of $L$ using the various time-derivatives of $f_{ext}(t)$; it is:

$$\left( \left( \ddot{x}_{torso} - \omega_{sp}^2 \ddot{x}_{torso} \right) - \omega_{st}^2 \left( \ddot{x}_{torso} - \omega_{sp}^2 x_{torso} \right) \right) \\ + m^{-1} \left( \ddot{f}_{ext} - \omega_{st}^2 f_{ext} \right) = 0. \tag{28}$$

The system of two differential equations given in (27) and (28) is solved by trajectories $x_{torso}(t)$ and $f_{ext}(t)$. The trajectory $x_{torso}(t)$ that solves the system is:

$$x_{torso}(t) = c_1 \sinh(\omega_{sp} t) + c_2 \cosh(\omega_{sp} t) \\ + c_3 t \sinh(\omega_{sp} t) + c_4 t \cosh(\omega_{sp} t) \\ + c_5 \sinh(\omega_{st} t) + c_6 \cosh(\omega_{st} t). \tag{29}$$

The trajectory $x_{torso}(t)$ of the torso consists of four terms in $\omega_{sp} t$ related to the motion of the stance leg and torso as an inverted pendulum acting under the force of gravity and two terms in $\omega_{sw} t$ related to the force of muscle acting on the torso. The four terms in $\omega_{sp} t$ taken together are the solution to a fourth-order differential equation corresponding to the solution of an optimal control problem in which the acceleration $\ddot{x}_{torso}(t)$ is controlled. The additional two terms in $\omega_{st} t$ provide a description of how the acceleration $\ddot{x}_{torso}(t)$ is in turn controlled by the forces generated by muscles. The trajectory $f_{ext}(t)$ of the external force that solves the system is:

$$f_{ext}(t) = k_1 \sinh(\omega_{st} t) + k_2 \cosh(\omega_{st} t). \tag{30}$$

The trajectories of the torso and the external force are parameterized by the eight parameters $c_1, \ldots, c_6$, and $k_1$ and $k_2$ in (29) and (30). We determine these parameter values by specifying the initial and final conditions the trajectories must satisfy.

The torso must traverse a step length $s$ in the avg. step time $T$. It must begin each step with some velocity $V_{torso}^- \geq 0$ and acceleration $A_{torso}^-$, and end each step with some velocity $V_{torso}^+ \geq 0$ and acceleration $A_{torso}^+$; that is:

$$\begin{aligned}
x_{torso}(-T/2) &= -s/2, & x_{torso}(T/2) &= s/2, \\
\dot{x}_{torso}(-T/2) &= V_{torso}^-, & \dot{x}_{torso}(T/2) &= V_{torso}^+, \\
\ddot{x}_{torso}(-T/2) &= A_{torso}^-, & \ddot{x}_{torso}(T/2) &= A_{torso}^+.
\end{aligned} \quad (31)$$

Combining (29) and (31) givens a linear system of six equations that allows us to solve the six parameters $c_1, \ldots, c_6$ in (29) in terms of the five parameters $s$, $V_{torso}^-$, $V_{torso}^+$, $A_{torso}^-$, and $A_{torso}^+$ in (31) using conventional methods of linear algebra. Thus, we can describe the optimal trajectory $x_{torso}(t)$ of the torso entirely by specifying the step-length, and the velocities and accelerations immediately before and after heel-strike together with the step-time. We discuss the physical meaning of $V_{torso}^-$, $V_{torso}^+$, $A_{torso}^-$, and $A_{torso}^+$ in (31) further in Sec. 3.7.

The external force must begin each step with some force $F^-$, and end each step with some force $F^+$; that is:

$$f_{ext}(-T/2) = F^-, \quad f_{ext}(T/2) = F^+. \quad (32)$$

Since we have allowed for a discontinuity due to heel-strike we allow $F^- \neq F^+$. The two parameters $k_1$ and $k_2$ in (30) are determined entirely by the two parameters $F^-$ and $F^+$ in (32) together with the step-time $T$. Thus, we can describe the optimal trajectory $f_{ext}(t)$ of the external force entirely by specifying the external forces immediately before and after heel-strike together with the step-time. We also need to average the required external force $F_{ext}$

$$\langle f_{ext}(t) \rangle = F_{ext}. \quad (33)$$

Combining (15), (29), and (30), we find that the magnitude $|\vec{F}_{st}(t)|$ of the force generated on the torso by the stance leg is:

$$\begin{aligned}
|\vec{F}_{st}(t)| \approx {}& \omega_{sp} m \cdot (c_3 \cosh(\omega_{sp} t) + c_4 \sinh(\omega_{sp} t)) \\
&+ \left((\omega_{st}^2 - \omega_{sp}^2) m c_5 + k_1\right) \sinh(\omega_{st} t) \\
&+ \left((\omega_{st}^2 - \omega_{sp}^2) m c_6 + k_2\right) \cosh(\omega_{st} t).
\end{aligned} \quad (34)$$

Of the eight parameters $c_1, \ldots, c_6$, and $k_1$ and $k_2$ that determine the optimal trajectories $x_{torso}(t)$ of the torso (29) and $f_{ext}(t)$ of the external force (30), only six, $c_3, \ldots, c_6$, $k_1$, and $k_2$ affect the magnitude $|\vec{F}_{st}(t)|$ of the force generated on the torso by the stance leg. The remaining two, $c_1, c_2$, describe the proper state of the torso so that the walking gait can be carried out correctly using the muscle activations given in (34).

*3.4.2 Hamiltonian Method*

The Hamiltonian $H_{st}$ governing the trajectories of the torso and external force is given by:

$$H_{st} = P_x^T \dot{Q}_x + P_f^T \dot{Q}_f - L_{st}. \quad (35)$$

The generalized coordinates are $Q_x$ and $Q_f$ are:

$$Q_x^{\mathrm{T}} = \begin{bmatrix} x_{torso}, & \dot{x}_{torso}, & \ddot{x}_{torso} \end{bmatrix}, \tag{36}$$
$$Q_f = f_{ext}.$$

The controls are the jerk $\dddot{x}_{torso}(t)$ and the yank $\dot{f}_{ext}(t)$:

$$\begin{aligned} u_x(t) &= \dddot{x}_{torso}(t), \\ u_f(t) &= \dot{f}_{ext}(t). \end{aligned} \tag{37}$$

We calculate the generalized momentum vectors $P_x$ and $P_f$ and the Hamiltonian $H_{st}$ along the optimal trajectory of the torso in Appendix 1; the Hamiltonian $H_{st}$ and generalized energy $\Psi_{st}$ of the torso satisfy:

$$\begin{aligned} H_{st} &= \alpha_{st} \cdot \left( \dot{F}_{st}^2 - \omega_{st}^2 F_{st}^2 \right) \\ &+ 2\alpha_{st} \cdot \left( \left( \ddot{F}_{st} - \omega_{st}^2 \dot{F}_{st} \right) \dot{x}_{torso} - \left( \dddot{F}_{st} - \omega_{st}^2 F_{st} \right) \ddot{x}_{torso} \right) \\ &= \Psi_{st} \end{aligned} \tag{38}$$

When moving in an optimal trajectory between heel-strikes, the generalized energy $\Psi_{st}$ is an approximate constant of the motion of the torso and external force insofar as the approximation in (22) holds. Although we refrain from working out its exact form here, the constant generalized energy $\Psi_{st}$ is expressed in terms of the eight parameters $c_1$, ..., $c_6$, and $k_1$ and $k_2$ in (29) and (30) using (38).

*3.5 Optimal Trajectory of the Swing Foot*

We calculate the optimal trajectory for the motion of the swing foot under the torso by finding the trajectory that minimizes the cost functional $J_{sw}$ in (25) following the same procedure that we have used in Sec. 3.4.

*3.5.1 Lagrangian Method*

The Lagrangian $L_{sw}$ governing the motion of the swing foot is the integrand of the cost functional $J_{sw}$ in (25):

$$\begin{aligned} L_{sw} &= \alpha_{sw} \mu^2 \left( \dddot{x}_{foot} + \omega_{sp}^2 \left( \dot{x}_{foot} - v \right) \right)^2 \\ &+ \varepsilon_{sw} \mu^2 \left( \ddot{x}_{foot} + \omega_{sp}^2 \left( x_{foot} - vt \right) \right)^2. \end{aligned} \tag{39}$$

We obtain single equation of motion from the Lagrangian by taking the partial derivatives of $L$ using the various time-derivatives of $x_{foot}(t)$; it is:

$$\begin{aligned} & \left( \ddddot{\ddot{x}}_{foot} + \omega_{sp}^2 \ddddot{x}_{foot} \right) - \omega_{sw}^2 \left( \dddot{x}_{foot} + \omega_{sp}^2 \ddot{x}_{foot} \right) \\ & + \omega_{sp}^2 \left( \left( \ddddot{x}_{foot} + \omega_{sp}^2 \ddot{x}_{foot} \right) - \omega_{sw}^2 \left( \ddot{x}_{foot} + \omega_{sp}^2 \left( x_{foot} - vt \right) \right) \right) = 0. \end{aligned} \tag{40}$$

The trajectory $x_{foot}(t)$ that solves (40) is:

$$\begin{aligned} x_{foot}(t) &= vt + c_1 \sin(\omega_{sp} t) + c_2 \cos(\omega_{sp} t) \\ &+ c_3 t \sin(\omega_{sp} t) + c_4 t \cos(\omega_{sp} t) \\ &+ c_5 \sinh(\omega_{sw} t) + c_6 \cosh(\omega_{sw} t). \end{aligned} \tag{41}$$

The trajectory of the swing foot is parameterized by the six parameters $c_1$, ..., $c_6$, in (41). As was the case with $x_{torso}(t)$, we observe that the trajectory $x_{foot}(t)$ consists of four terms in $\omega_{sp}t$ related to the motion of the swing leg as a simple pendulum acting under the force of gravity and two terms in $\omega_{sw}t$ related to the force of muscle acting on the swing foot. We may again understand the two terms in $\omega_{sw}t$ as providing a description of how the acceleration $\ddot{x}_{foot}(t)$ is controlled by the muscles.

The swing foot must traverse an avg. stride length $2s$ in the avg. step time $T$. It must begin each step resting on the ground with no velocity or acceleration but may end each step with some velocity $V_{foot}^+ \geq 0$ and acceleration $A_{foot}^+$, assuming impact on the ground bring it to a complete stop for reasonable values of $V_{foot}^+$ and $A_{foot}^+$. The initial and final conditions of the motion of the swing foot for a single step are:

$$\begin{aligned}
x_{foot}(-T/2) &= -s, & x_{foot}(T/2) &= s, \\
\dot{x}_{foot}(-T/2) &= 0, & \dot{x}_{foot}(T/2) &= V_{foot}^+, \\
\ddot{x}_{foot}(-T/2) &= 0, & \ddot{x}_{foot}(T/2) &= A_{foot}^+.
\end{aligned} \qquad (42)$$

Combining (41) and (42) gives a linear system of six equations that allows us to solve the six parameters $c_1$, ..., $c_6$ in (41) in terms of the three parameters $s$, $V_{foot}^+$, and $A_{foot}^+$ in (42). Thus, we can describe the optimal trajectory $x_{foot}(t)$ of the swing foot entirely by specifying the step-length, and the velocity and acceleration immediately before heel-strike together with the step-time. We discuss the physical meaning of $V_{foot}^+$, and $A_{foot}^+$ in (42) further in Sec. 3.7.

Combining (18) and (41), we find that the magnitude $|\vec{F}_{sw}(t)|$ of the force generated on the swing foot by the swing leg is:

$$\begin{aligned}
|\vec{F}_{sw}(t)| \approx{} & \omega_{sp} \cdot (c_3 \cos(\omega_{sp}t) - c_4 \sin(\omega_{sp}t)) \\
& + (\omega_{sp}^2 + \omega_{sw}^2)\mu \cdot (c_5 \sinh(\omega_{st}t) + c_6 \cosh(\omega_{st}t)).
\end{aligned} \qquad (43)$$

We observe that of the six parameters $c_1$, ..., $c_6$, that determine the optimal trajectory $x_{foot}(t)$ of the swing foot in (41), four, $c_3$, ..., $c_6$, affect the magnitude $|\vec{F}_{sw}(t)|$ of the force generated on the swing foot by the swing leg. The remaining two, $c_1$ and $c_2$, describe the proper state of the swing foot so that the walking gait can be carried out correctly using the muscle activations given in (43).

*3.5.2 Hamiltonian Method*

The Hamiltonian $H_{sw}$ governing the trajectory of the swing foot is given by:

$$H_{sw} = P^{\mathrm{T}}\dot{Q} - L_{sw}. \qquad (44)$$

The generalized coordinates vector Q is:

$$Q^{\mathrm{T}} = \begin{bmatrix} x_{foot}, & \dot{x}_{foot}, & \ddot{x}_{foot} \end{bmatrix}. \qquad (45)$$

Thus, the generalized coordinates vector Q contains the position $x_{foot}(t)$, velocity $\dot{x}_{foot}(t)$, and acceleration $\ddot{x}_{foot}(t)$. The control is the jerk $\dddot{x}_{foot}(t)$:

$$u(t) = \dddot{x}_{foot}(t). \qquad (46)$$

We calculate the generalized momentum vector $P$ and the Hamiltonian $H_{sw}$ along the optimal trajectory of the swing foot in Appendix 2; the Hamiltonian $H_{sw}$ and generalized energy $\Psi_{sw}$ in this case satisfy:

$$\begin{aligned} H_{sw} &= \alpha_{sw} \cdot \left( \dot{F}_{sw}^2 - \omega_{sw}^2 F_{sw}^2 \right) \\ &\quad + 2\alpha_{sw} \cdot \left( \left( \ddot{F}_{sw} - \omega_{sw}^2 \dot{F}_{sw} \right) \dot{x}_{foot} - \left( \ddot{F}_{sw} - \omega_{sw}^2 F_{sw} \right) \ddot{x}_{foot} \right) \\ &= \Psi_{sw}. \end{aligned} \qquad (47)$$

When moving in an optimal trajectory between heel-strikes, the generalized energy $\Psi_{sw}$ is an approximate constant of the motion of the swing foot insofar as the approximation in (22) holds. Although we refrain from working out its exact form here, the constant generalized energy $\Psi_{sw}$ is expressed in terms of the six parameters $c_1, \ldots, c_6$, in (41) using (47).

*3.6 Mechanical Energy of Walking Gait*

The kinetic mechanical energy $U_K(t)$ of a walking gait can be calculated from the trajectories of the segments using the usual definition of kinetic energy from classical mechanics; it is:

$$U_K = (1/2) m \dot{x}_{torso}^2 + (1/2) \mu \dot{x}_{foot}^2. \qquad (48)$$

The kinetic mechanical energy $U_K(t)$ in (48) changes over time reflecting an ongoing process of changes in mechanical energy to walking gait. We will assume that, whenever one of the torso or swing foot gains kinetic mechanical energy while the other loses it, the body transfers kinetic mechanical energy between the two. Kinetic mechanical energy may change for two reasons: (i) interaction with the environment, or (ii) action of the muscles on the segments of the body. If we assume that the only interaction with the environment that happens is a change in kinetic mechanical energy at heel-strike, then any other change in kinetic mechanical energy must be due the action of muscles.

The change in kinetic mechanical energy at heel-strike is given by the values of the parameters $V_{torso}^+$ and $V_{foot}^+$ just before heel-strike and $V_{torso}^-$ just after heel-strike. For the case where $V_{torso}^- < V_{torso}^+$, the torso has lost some amount of kinetic mechanical energy at heel-strike while for the case where $V_{torso}^- > V_{torso}^+$, the torso has gained some amount of kinetic mechanical energy. For the case where $V_{foot}^+ > 0$, the swing foot has lost some amount of kinetic mechanical energy at heel-strike. As we have assumed that, whenever possible, the body transfers kinetic mechanical energy between the torso and the swing foot, the change in kinetic mechanical energy $\Delta U_K(t)$ at heel-strike is:

$$\Delta U_K = (1/2) m \cdot \left[ \left( V_{torso}^- \right)^2 - \left( V_{torso}^+ \right)^2 \right] - (1/2) \mu \cdot \left( V_{foot}^+ \right)^2. \qquad (49)$$

The muscles must act during the following step to restore kinetic mechanical energy to the body so that this change in mechanical energy can occur at the next heel-strike. The optimal trajectories $x_{torso}(t)$ and $x_{foot}(t)$ of the torso and swing foot, respectively, describe the effect of this action of by the muscles on scales below a step.

Using (48), we find the rate of change of the kinetic mechanical energy over the course of the walking gait cycle is:

$$\dot{U}_K = m \dot{x}_{torso} \ddot{x}_{torso} + \mu \dot{x}_{foot} \ddot{x}_{foot}. \qquad (50)$$

The action of muscles accrues a cost in metabolic energy we can associate a metabolic rate with (50). Thus, we should be able to relate (50) to a metabolic rate for intervals between heel-strikes. Rather than provide a detailed analysis of how the muscles act on the torso and the swing foot separately, we follow (2) and simply assume that we can make an approximation $\dot{W} \approx \eta^{+/-} \dot{U}_K$ where $\eta^{+/-}$ may take on one value when $\dot{U}_K > 0$ and another when $\dot{U}_K < 0$ so that $\dot{W} \geq 0$. Thus, the metabolic rate can differ between the cases where mechanical energy is added to or removed from the body. Therefore, metabolic rate associated with the muscles mediating the change in kinetic mechanical energy is:

$$\dot{W} \approx \eta^{+/-} \cdot \left( m\dot{x}_{torso}\ddot{x}_{torso} + \mu\dot{x}_{foot}\ddot{x}_{foot} \right). \tag{51}$$

Integrating over a step, (51) provides an additional metabolic energy not explicitly contained in the model used in [14-16], but implicitly contained in it in the form of an additive constant that was introduced to facilitate fitting the model to the empirical data in Atzler & Herbst. [17]

*3.7 Discussion*

When constructing the general form of a cost functional for a movement in (5), we required that there be no discontinuities in the force. Nevertheless, the model of walking gait we use precisely does have such discontinuities at heel-strike. In fact, there are two distinct discontinuities happening simultaneously in our model at heel-strike: (i) the impact of the previous swing foot on the ground and (ii) the lifting of the new swing foot from the ground. Thus, we can argue that, at the scale at which we are working, while there might be no discontinuities in the muscle forces at heel-strike, there are discontinuities related to the discontinuous interaction between the body and the ground.

Between heel-strikes, the trajectories of the torso and swing foot are completely determined by the parameter values $V_{torso}^-$, $V_{torso}^+$, $V_{foot}^+$, $A_{torso}^-$, $A_{torso}^+$, and $A_{foot}^+$, and $F^-$ and $F^+$ in (31), (32), and (42). At heel-strike, these parameter values summarize any jarring of the body by the impact of the swing foot with the ground. For the intervals of time between heel-strikes these parameters serve to completely determine the trajectories of the torso and swing foot as the body recovers from the previous heel-strike and prepares for the next. In effect, we can look at the optimal control model of walking gait that we have developed as one in which the subject controls each heel-strike to produce the initial and final conditions required to generate the desired trajectories of the torso and swing foot between heel-strikes. In this respect, the optimal control model is a kind of generalization of the passive-dynamic walking model [26] in which a simple walking mechanism steps down a slightly inclined plane with the mechanism moving in a physically determined way under the influence of gravity and with the impact at heel-strike driving the mechanism into the next step while gravity provides for the restoration of lost kinetic mechanical energy.

While the optimal control model of walking begins to develop a more detailed description of walking gait on scale below that of a step, it leaves out the details of many aspects of walking gait such the double support portion of walking gait where both feet are on the ground, the thrust on the torso at toe off, the lifting of the swing foot up from the ground, and any vertical motion of the torso. One of the advantages of using a variational approach to modeling movement trajectories, both in physics and in the present study of the biomechanics of human movement, is the ease with which existing models can be modified to produce more detailed models by the introduction of additional terms in the cost functional. Thus, the relatively simple optimal control model of walking gait that we construct here may be readily modified to accommodate more aspects of walking gait, such as the one we have listed, by expanding the cost functional we have given with additional terms that describe further details of walking gait. Work in this direction might begin using the simple walking gait models in [27, 28] that include descriptions of such aspects of walking gait.

## 4 Symmetric & Uniform Walking Gait in the Low Yank Limit

We now show that the optimal control model in Sec. 3 generalizes the model in [14-16] so that we can look at the optimal control model as a means of generalizing the model in [14-16] to scales below that of a step. While it may not be surprising that the present optimal control model is able to approximate the model in [14-16], it is of interest to show the conditions under which it does so. We find that we arrive at the approximation by solving the optimal control model for the case of walking gaits that are symmetric and uniform and looking at walking gaits in which the yanks are kept relatively low.

*4.1 Symmetric Walking Gait*

We obtain a symmetric trajectory for the torso during a step by requiring it have the same velocity at the beginning and end of a step, and the same magnitude of acceleration at the beginning and end of a step. Thus, we restrict the parameter values $V_{torso}^-$, $V_{torso}^+$, $A_{torso}^-$, and $A_{torso}^+$ in (31) so that $V_{torso}^- = V_{torso}^+$, and $|A_{torso}^-| = |A_{torso}^+|$, and we replace (31) with:

$$\begin{aligned}
x_{torso}\left(-T/2\right) &= -s/2, & x_{torso}\left(T/2\right) &= s/2, \\
\dot{x}_{torso}\left(-T/2\right) &= V_{torso}, & \dot{x}_{torso}\left(T/2\right) &= V_{torso}, \\
\ddot{x}_{torso}\left(-T/2\right) &= -A_{torso}, & \ddot{x}_{torso}\left(T/2\right) &= A_{torso}.
\end{aligned} \qquad (52)$$

Similarly, we obtain a symmetric external force during a step by requiring that the external force to be the same at the beginning and end of a step. Thus, we restrict the parameter values $F^-$ and $F^+$ in (32) so that $F^- = F^+$, and we replace (32) with:

$$f_{ext}\left(-T/2\right) = F, \quad f_{ext}\left(T/2\right) = F. \qquad (53)$$

Finally, we obtain a symmetric trajectory for the swing foot during a step by requiring that it have the same velocity at the beginning and end of a step, and the same magnitude of acceleration at the beginning and end of a step. This can only happen when we take $V_{torso}^+ = 0$, and $A_{foot}^+ = 0$; thus we replace (42) with:

$$\begin{aligned}
x_{foot}\left(-T/2\right) &= -s, & x_{foot}\left(T/2\right) &= s, \\
\dot{x}_{foot}\left(-T/2\right) &= 0, & \dot{x}_{foot}\left(T/2\right) &= 0, \\
\ddot{x}_{foot}\left(-T/2\right) &= 0, & \ddot{x}_{foot}\left(T/2\right) &= 0.
\end{aligned} \qquad (54)$$

*4.2 Uniform Walking Gait*

A walking gait is approximately uniform when the torso moves with an approximately constant speed:

$$x_{torso}\left(t\right) \approx vt. \qquad (55)$$

*4.3 Optimal Trajectories of Symmetric and Uniform Walking Gait*

In the case of symmetric walking gait, only three of the six terms in the optimal trajectory $x_{torso}(t)$ of the torso in (29) remain (have non-zero parameter values), namely those that are odd in time on the interval $-T/2 \leq t \leq T/2$. This gives an optimal trajectory $x_{torso}(t)$ of the torso of the form:

$$x_{torso}(t) = C_1 \sinh(\omega_{sp} t) + C_2 t \cosh(\omega_{sp} t) + C_3 \sinh(\omega_{st} t). \tag{56}$$

In Appendix 3, we calculate the values of the parameters $C_1$, $C_2$, and $C_3$ that give a trajectory $x_{torso}(t)$ of the torso that approximates uniformity as given in (55); these are:

$$\begin{aligned} C_1 &\approx -(\omega_{st}^3 / \omega_{sp}^5)(\omega_{sp}^2 + (3/2)(\omega_{sp}^2 - \omega_{st}^2))C_3, \\ C_2 &\approx (1/2)(\omega_{st}^3 / \omega_{sp}^4)(\omega_{sp}^2 - \omega_{st}^2)C_3, \\ C_3 &\approx \left( \frac{\omega_{sp}^4}{\omega_{st}(\omega_{sp}^2 - \omega_{st}^2)^2} \right) v. \end{aligned} \tag{57}$$

For symmetric walking gait, only one of the two terms in the optimal trajectory $f_{ext}(t)$ of the external force in (29) remains (has non-zero parameter value), namely the one that is even in time on the interval $-T/2 \leq t \leq T/2$:

$$f_{ext}(t) = K \cosh(\omega_{st} t). \tag{58}$$

In Appendix 3, we calculate the value of the parameter $K$ that gives a trajectory $f_{ext}(t)$ of the external force that has symmetry as given in (53); these are:

$$K = \left[ 1 + \left( \frac{\sinh(\omega_{st} T / 2)}{\omega_{st} T / 2} - 1 \right) \right]^{-1} F_{ext}. \tag{59}$$

For symmetric walking gait, only three of the six terms in the optimal trajectory $x_{foot}(t)$ of the swing foot in (41) only three remain (have non-zero parameter values), namely those that are odd in time on the interval $-T/2 \leq t \leq T/2$. This gives an optimal trajectory $x_{foot}(t)$ of the swing foot of the form:

$$x_{foot}(t) = vt + C_1 \sin(\omega_{sp} t) + C_2 t \cos(\omega_{sp} t) + C_3 \sinh(\omega_{sw} t). \tag{60}$$

In constructing the walking gait model in Sec. 3, we assumed small $\omega_{sp} t$. We may therefore make the approximation $\sin(\omega_{sp} t) \approx \omega_{sp} t - (\omega_{sp} t)^3/6$ allowing us to approximate (60) as:

$$x_{foot}(t) \approx vt + \kappa_1 t + \kappa_2 t^3 + C_3 \sinh(\omega_{sw} t). \tag{61}$$

In Appendix 4, we calculate the values of the parameters $\kappa_1$, $\kappa_2$, and $C_3$ that give a trajectory $x_{foot}(t)$ that has symmetry as given in (54); these are:

$$\kappa_1 = 2v + \left[\frac{(\omega_{sw}T)^2 - 24}{12T}\right]\sinh(\omega_{sw}T/2)C_3,$$

$$\kappa_2 = -\left[\frac{(\omega_{sw}T)^2}{3T^3}\right]\sinh(\omega_{sw}T/2)C_3, \tag{62}$$

$$C_3 = \left[\frac{12T}{\left(12 + (\omega_{sw}T)^2\right)\sinh(\omega_{sw}T/2) - 6(\omega_{sw}T)\cosh(\omega_{sw}T)}\right]v.$$

*4.4 Metabolic Energy in the Low Yank Limit*

We limit the model to low yanks by requiring the model parameters $\alpha_{st}$ and $\varepsilon_{st}$ in (24), and $\alpha_{sw}$ and $\varepsilon_{sw}$ in (25) to have values such that for typical walking gaits we have $\alpha_{st}\langle \dot{F}_{st}^2\rangle >> \varepsilon_{st}\langle F_{st}^2\rangle$ and $\alpha_{sw}\langle \dot{F}_{sw}^2\rangle >> \varepsilon_{sw}\langle F_{sw}^2\rangle$. This has the effect of making the yank terms contribute most of the cost in the cost functional in (19) so that optimal trajectories will tend to have lower yanks. In practice, we take the low yank limit by taking the limit as $\omega_{st}$ and $\omega_{sw}$ in (20) become relatively low frequencies. As the metabolic energy model used in [14-16] was a model of the metabolic energy per step of walking gait, we look specifically at the metabolic energy per step in the low yank limit.

Following (3), the metabolic energy per step of the torso is:

$$W_{torso} = \int_{-T/2}^{T/2} \left(\varepsilon_{st} F_{st}^2 + \eta_{ext} \vec{f}_{ext} \cdot \dot{\vec{x}}_{torso}\right) dt. \tag{63}$$

We can split (63) into the sum $W_{torso} = W_{st} + W_{ext}$ where $W_{st}$ is metabolic energy per step associated with generating force on the torso by the stance leg and $W_{ext}$ is the metabolic energy per step associated with doing external work. The metabolic energy per step $W_{st}$ associated with generating force on the torso is:

$$W_{st} = \varepsilon_{st} \int_{-T/2}^{T/2} F_{st}^2 dt. \tag{64}$$

The metabolic energy per step $W_{ext}$ associated with doing external work is:

$$W_{ext} = \eta_{ext} \int_{-T/2}^{T/2} \vec{f}_{ext} \cdot \dot{\vec{x}}_{torso} dt. \tag{65}$$

We first look the metabolic energy per step $W_{st}$ associated generating force on the torso by the stance leg in (64). Combining (64) with (56) we obtain:

$$\begin{aligned}W_{st} \approx \varepsilon_{st}\int_{-T/2}^{T/2} &\left(\omega_{sp}mC_2\sinh(\omega_{sp}t)\right.\\ &+\left(\omega_{st}^2 - \omega_{sp}^2\right)mC_3\sinh(\omega_{st}t)\\ &+\left.K\cosh(\omega_{st}t)\right)^2 dt.\end{aligned} \tag{66}$$

The constants in (66) are given in (57) and (59). In Appendix 5, we take low yank limit by taking $\omega_{st} \to 0$, we find:

$$W_{st} \approx \left(\varepsilon_{st} g^2 m^2 / 12 L^2\right) s^3 / v + \left(\varepsilon_{st}\right) F_{ext}^2 s / v. \qquad (67)$$

In this same limit, we also find:

$$f_{ext} \approx F_{ext}. \qquad (68)$$

The metabolic energy $W_{ext}$ associated with doing external work in (65) therefore becomes:

$$W_{ext} \approx \eta_{ext} F_{ext} s. \qquad (69)$$

Evaluating $W_{torso}$ by combining (67) and (69) using the sum $W_{torso} = W_{st} + W_{ext}$ we obtain the model of walking gait doing external work used in [15, 16], as we should.

Following (3), the metabolic energy per step of the swing foot is:

$$W_{foot} = \varepsilon_{sw} \int_{-T/2}^{T/2} F_{sw}^2 dt. \qquad (70)$$

Combining (70) with (66), the metabolic energy per step $W_{foot}$ associated with the motion of the swing foot is:

$$W_{foot} \approx \varepsilon_{sw} \mu^2 \int_{-T/2}^{T/2} \left(6\kappa_2 t + C_3 \omega_{sw}^2 \sinh\left(\omega_{sw} t\right)\right)^2 dt. \qquad (71)$$

In Appendix 6, we take low yank limit by taking $\omega_{sw} \to 0$, we find:

$$W_{foot} \approx 192 \varepsilon_{sw} \mu^2 v^3 / s. \qquad (72)$$

This bears a functional resemblance to the swing foot trajectory used in [14-16]. We expect that it can approximate the model used in [14-16] reasonably well over the range of empirical data used in validating that model.

*4.5 Discussion*

In the symmetric and uniform walking gait model, the torso maintains a constant kinetic mechanical energy while the swing foot gains kinetic mechanical energy up to midswing and then loses mechanical energy from then until heel-strike. The swing foot slows to zero velocity relative to the ground and there is no jarring impact with the ground. Thus, all change in the kinetic mechanical energy of the body is affected by the action of the muscles. However, it is clear from our discussion of heel-strike in Sec. 3 that we do not expect walking gaits to be symmetric, but instead we expect there to be a jarring due to impact with the ground. We should therefore begin the process of using the optimal control model of walking gait to generalize the walking gait model in [14-16] to more realistic asymmetric walking gaits.

**Appendix 1**

We would like to calculate the Hamiltonian $H_{st}$ of the torso as it is moved by the forces generated by the stance leg and generates the external force. The Hamiltonian $H_{st}$ satisfies (35); that is:

$$H_{st} = P_x^{\mathrm{T}} \dot{Q}_x + P_f^{\mathrm{T}} \dot{Q}_f - L_{st}. \tag{73}$$

The Hamiltonian $H_{st}$ consists of a term $P_x^T \dot{Q}_x$ associated with the trajectory of the torso and $P_f^T \dot{Q}_f$ associated with the trajectory of the external force, and the Lagrangian $L_{st}$ in (26) The generalized coordinates vectors $Q_x$ and $Q_f$ given in (36).

We first calculate the quantity $P_x^T \dot{Q}_x$ associated with the trajectory of the torso. The generalized momentum vector $P_x$ and the control $u_x$ in (37) are:

$$\begin{aligned} P_x^{\mathrm{T}} &= \begin{bmatrix} p_1, & p_2, & p_3 \end{bmatrix}, \\ u_x &= \dddot{x}_{torso}. \end{aligned} \tag{74}$$

Using equation C in (11), we find the optimal trajectory satisfies $\partial H_{st}/\partial u_x = 0$; this gives:

$$\begin{aligned} p_3 &= 2\alpha_{st} m^2 \left( u_x - \omega_{sp}^2 \dot{x}_{torso} + \dot{f}_{ext}/m \right) \\ &= 2\alpha_{st} m \dot{F}_{st}. \end{aligned} \tag{75}$$

Using equation B in (11), we find that the generalized momentum vector $P_x$ for optimal trajectory satisfies $\dot{P}_x = -\partial H_{st}/\partial Q_x$; this gives:

$$\dot{P}_x = -\begin{bmatrix} 0 & 0 & 0 \\ 1 & 0 & 0 \\ 0 & 1 & 0 \end{bmatrix} P_x + 2\alpha_{st} m \begin{bmatrix} -\omega_{st}^2 \omega_{sp}^2 F_{st} \\ -\omega_{sp}^2 \dot{F}_{st} \\ \omega_{st}^2 F_{st} \end{bmatrix}. \tag{76}$$

Therefore, the generalized momentum vector $P_x$ is:

$$P_x = 2\alpha_{st} m \begin{bmatrix} -\omega_{sp}^2 \\ 0 \\ 1 \end{bmatrix} \dot{F}_{st} + 2\alpha_{st} m \begin{bmatrix} \left(\dddot{F}_{st} - \omega_{st}^2 \dot{F}_{st}\right) \\ -\left(\ddot{F}_{st} - \omega_{st}^2 F_{st}\right) \\ 0 \end{bmatrix}. \tag{77}$$

We finally find that $P_x^T \dot{Q}_x$ is:

$$\begin{aligned} P_x^{\mathrm{T}} \dot{Q}_x &= 2\alpha_{st} \dot{F}_{st} \cdot \left( m \dddot{x}_{torso} - \omega_{sp}^2 m \dot{x}_{torso} \right) \\ &\quad + 2\alpha_{st} \cdot \left( \left( \dddot{F}_{st} - \omega_{st}^2 \dot{F}_{st} \right) \dot{x}_{torso} - \left( \ddot{F}_{st} - \omega_{st}^2 F_{st} \right) \ddot{x}_{torso} \right). \end{aligned} \tag{78}$$

We next calculate the quantity $P_f^T \dot{Q}_f$ associated with the trajectory of the external force. The generalized momentum vector $P_f$ and the control $u_f$ in (37) are:

$$\begin{aligned} P_f^{\mathrm{T}} &= p, \\ u_f &= \dot{f}_{ext}. \end{aligned} \tag{79}$$

Using equation C in (11), we find optimal trajectory satisfies $\partial H_{st}/\partial u_f = 0$; this gives:

$$\begin{aligned} p &= 2\alpha_{st} m \left( \dddot{x}_{torso} - \omega_{sp}^2 \dot{x}_{torso} + u_f/m \right) \\ &= 2\alpha_{st} \dot{F}_{st}. \end{aligned} \tag{80}$$

The generalized momentum vector $P_f$ (see Appendix 1) is:

$$P_f = 2\alpha_{st} \dot{F}_{st}. \tag{81}$$

We finally find that $P_f^T \dot{Q}_f$ is:

$$P_f^T \dot{Q}_f = 2\alpha_{st} \dot{F}_{st} \dot{f}_{ext}. \tag{82}$$

Combining (26), (36), (73), (77), and (81), we find that the Hamiltonian $H_{st}$ governing the motion of the torso is:

$$\begin{aligned} H_{st} &= \alpha_{st} \cdot \left( \dot{F}_{st}^2 - \omega_{st}^2 F_{st}^2 \right) \\ &+ 2\alpha_{st} \cdot \left( \left( \ddot{F}_{st} - \omega_{st}^2 \dot{F}_{st} \right) \dot{x}_{torso} - \left( \dot{F}_{st} - \omega_{st}^2 F_{st} \right) \ddot{x}_{torso} \right) \\ &= \Psi_{st} \end{aligned} \tag{83}$$

Thus, the generalized energy $\Psi_{st}$ is an approximate constant of the motion of the torso that holds according to how well the approximation in (22) holds. When moving in an optimal trajectory, the torso moves over the course of a step, between heel-strikes, in a way that keeps the Hamiltonian $H_{st}$ approximately constant according to (83).

**Appendix 2**

We would like to calculate the Hamiltonian $H_{sw}$ of the swing foot as it is moved by the forces generated by the swing leg. The Hamiltonian $H_{sw}$ satisfies (44); that is:

$$H_{sw} = P^T \dot{Q} - L_{sw}. \tag{84}$$

The Hamiltonian $H_{sw}$ consists of a term $P^T \dot{Q}$ associated with the trajectory of the swing foot, and the Lagrangian $L_{sw}$ in (39). The generalized coordinates vector $Q$ given in (45).

We calculate the quantity $P^T \dot{Q}$ associated with the trajectory of the swing foot. The generalized momentum vector $P$ and the control $u$ are:

$$\begin{aligned} P^T &= \begin{bmatrix} p_1, & p_2, & p_3 \end{bmatrix}, \\ u &= \ddot{x}_{foot}. \end{aligned} \tag{85}$$

Using equation C in (11), we find optimal trajectory satisfies $\partial H_{sw}/\partial u = 0$; this gives:

$$\begin{aligned} p_3 &= 2\alpha_{sw} \mu^2 \left( u + \omega_{sp}^2 \dot{x}_{foot} \right) \\ &= 2\alpha_{sw} \mu \dot{F}_{sw}. \end{aligned} \tag{86}$$

Using equation B in (11), we find that the generalized momentum $P$ for optimal trajectory satisfies $\dot{P} = -\partial H_{sw}/\partial Q$; this gives:

$$\dot{P} = -\begin{bmatrix} 0 & 0 & 0 \\ 1 & 0 & 0 \\ 0 & 1 & 0 \end{bmatrix} P + 2\alpha_{sw}\mu \begin{bmatrix} \omega_{sw}^2 \omega_{sp}^2 F_{sw} \\ \omega_{sp}^2 \dot{F}_{sw} \\ \omega_{sw}^2 F_{sw} \end{bmatrix}. \tag{87}$$

The generalized momentum vector $P$ is:

$$P = 2\alpha_{sw} \begin{bmatrix} \omega_{sp}^2 \\ 0 \\ 1 \end{bmatrix} \dot{F}_{sw} + 2\alpha_{sw} \begin{bmatrix} \left(\dddot{F}_{sw} - \omega_{sw}^2 \dot{F}_{sw}\right) \\ -\left(\ddot{F}_{sw} - \omega_{sw}^2 F_{sw}\right) \\ 0 \end{bmatrix}. \tag{88}$$

We finally find that $P^T \dot{Q}$ is:

$$\begin{aligned} P^{\mathrm{T}}\dot{Q} &= 2\alpha_{sw}\dot{F}_{sw}^2 \\ &+ 2\alpha_{sw} \cdot \left(\left(\dddot{F}_{sw} - \omega_{sw}^2 \dot{F}_{sw}\right)\dot{x}_{foot} - \left(\ddot{F}_{sw} - \omega_{sw}^2 F_{sw}\right)\ddot{x}_{foot}\right). \end{aligned} \tag{89}$$

Combining (39), (45), (84), and (88), we find that the Hamiltonian $H_{sw}$ is:

$$\begin{aligned} H_{sw} &= \alpha_{sw} \cdot \left(\dot{F}_{sw}^2 - \omega_{sw}^2 F_{sw}^2\right) \\ &+ 2\alpha_{sw} \cdot \left(\left(\dddot{F}_{sw} - \omega_{sw}^2 \dot{F}_{sw}\right)\dot{x}_{foot} - \left(\ddot{F}_{sw} - \omega_{sw}^2 F_{sw}\right)\ddot{x}_{foot}\right) \\ &= \Psi_{sw}. \end{aligned} \tag{90}$$

Thus, the generalized energy $\Psi_{sw}$ is an approximate constant of the motion of the swing foot that holds according to how well the approximation in (22) holds. When moving in an optimal trajectory, the swing foot moves over the course of a step, between heel-strikes, in a way that keeps the Hamiltonian $H_{sw}$ approximately constant according to (47).

**Appendix 3**

The optimal trajectory $x_{torso}(t)$ of the torso for symmetric walking gait for a step taking place on the time interval $-T/2 \leq t \leq T/2$ has the general form given in (56); that is:

$$x_{torso}(t) = C_1 \sinh(\omega_{sp} t) + C_2 t \cosh(\omega_{sp} t) + C_3 \sinh(\omega_{st} t). \tag{91}$$

We would like to find parameters $C_1$, $C_2$, and $C_3$ so that the trajectory $x_{torso}(t)$ of the torso is approximately that of uniform walking gait as given by (55); namely:

$$x_{torso}(t) \approx vt. \tag{92}$$

We expand the symmetric torso trajectory in (91) as a power series in $t$ using the usual series expansions of the sinh and cosh functions:

$$\begin{aligned} x_{torso}(t) &= C_1 \sum_{n=0}^{\infty} \frac{(\omega_{sp} t)^{2n+1}}{(2n+1)!} + C_2 t \sum_{n=0}^{\infty} \frac{(\omega_{sp} t)^{2n}}{(2n)!} \\ &+ C_3 \sum_{n=0}^{\infty} \frac{(\omega_{st} t)^{2n+1}}{(2n+1)!}. \end{aligned} \tag{93}$$

As we have three parameter values $C_1$, $C_2$, and $C_3$ to solve for, we would like to truncate the series expansion in (93) to three independent terms. We do this by truncating the expansion to fifth-order in $t$; we find:

$$x_{torso}(t) \approx \left(C_1 \omega_{sp} + C_2 + C_3 \omega_{sp}\right)t \\
+ \left(C_1 \omega_{sp}^3 + 3C_2 \omega_{sp}^2 + C_3 \omega_{st}^3\right)t^3 / 3! \\
+ \left(C_1 \omega_{sp}^5 + 5C_2 \omega_{sp}^4 + C_3 \omega_{st}^5\right)t^5 / 5!. \tag{94}$$

Combining (92) and (94), we find that we can calculate $C_1$, $C_2$, and $C_3$ by solving a linear system of three equations in the three unknowns $C_1$, $C_2$, and $C_3$:

$$\begin{bmatrix} \omega_{sp} & 1 & \omega_{st} \\ \omega_{sp}^3 & 3\omega_{sp}^2 & \omega_{st}^3 \\ \omega_{sp}^5 & 5\omega_{sp}^4 & \omega_{st}^5 \end{bmatrix} \begin{bmatrix} C_1 \\ C_2 \\ C_3 \end{bmatrix} \approx \begin{bmatrix} v \\ 0 \\ 0 \end{bmatrix}. \tag{95}$$

The parameter values $C_1$, $C_2$, and $C_3$ that satisfy (92) are:

$$C_1 \approx -\left(\omega_{st}^3 / \omega_{sp}^5\right)\left(\omega_{sp}^2 + (3/2)\left(\omega_{sp}^2 - \omega_{st}^2\right)\right) C_3, \\
C_2 \approx (1/2)\left(\omega_{st}^3 / \omega_{sp}^4\right)\left(\omega_{sp}^2 - \omega_{st}^2\right) C_3, \\
C_3 \approx \left(\frac{\omega_{sp}^4}{\omega_{st}\left(\omega_{sp}^2 - \omega_{st}^2\right)^2}\right) v. \tag{96}$$

**Appendix 4**

The optimal trajectory $x_{foot}(t)$ of the swing foot for symmetric walking gait for a step taking place on the time interval $-T/2 \le t \le T/2$ has the general form given in (61); that is:

$$x_{foot}(t) \approx vt + \kappa_1 t + \kappa_2 t^3 + C_3 \sinh(\omega_{sw} t). \tag{97}$$

We would like to find a swing foot trajectory $x_{foot}(t)$ that satisfies the initial and final conditions given in (54); namely:

$$x_{foot}(-T/2) = -s, \quad x_{foot}(T/2) = s, \\
\dot{x}_{foot}(-T/2) = 0, \quad \dot{x}_{foot}(T/2) = 0, \\
\ddot{x}_{foot}(-T/2) = 0, \quad \ddot{x}_{foot}(T/2) = 0. \tag{98}$$

We can make the symmetric swing foot trajectory in (97) satisfy (98) by finding the three parameter values $\kappa_1$, $\kappa_2$, and $C_3$ that satisfy the linear system of three equations given by:

$$\begin{bmatrix} 4T & T^3 & 8\sinh(\omega_{sw} T / 2) \\ 4 & 3T^2 & 4\omega_{sw} \cosh(\omega_{sw} T / 2) \\ 0 & 3T & \omega_{sw}^2 \sinh(\omega_{sw} T / 2) \end{bmatrix} \begin{bmatrix} \kappa_1 \\ \kappa_2 \\ C_3 \end{bmatrix} = \begin{bmatrix} 8s \\ 0 \\ 0 \end{bmatrix}. \tag{99}$$

The parameter values $\kappa_1$, $\kappa_2$, and $C_3$ that satisfy (99) are:

$$\kappa_1 \approx 2v + \left(\frac{(\omega_{sw}T)^2 - 24}{12T}\right)\sinh(\omega_{sw}T/2)C_3,$$

$$\kappa_2 \approx -\left(\frac{(\omega_{sw}T)^2}{3T^3}\right)\sinh(\omega_{sw}T/2)C_3, \tag{100}$$

$$C_3 \approx \left(\frac{12T}{\left(12 + (\omega_{sw}T)^2\right)\sinh(\omega_{sw}T/2) - 6(\omega_{sw}T)\cosh(\omega_{sw}t)}\right)v.$$

**Appendix 5**

The metabolic energy per step is associated with the motion of the torso is given in (66); it is:

$$W_{st} \approx \varepsilon_{st}\int_{-T/2}^{T/2}\big(\omega_{sp}mC_2\sinh(\omega_{sp}t)$$
$$+ (\omega_{st}^2 - \omega_{sp}^2)mC_3\sinh(\omega_{st}t) \tag{101}$$
$$+ K\cosh(\omega_{st}t)\big)^2 dt.$$

The constants are:

$$C_2 \approx (1/2)\big(\omega_{st}^3/\omega_{sp}^4\big)\big(\omega_{sp}^2 - \omega_{st}^2\big)C_3,$$

$$C_3 \approx \left(\frac{\omega_{sp}^4}{\omega_{st}\big(\omega_{sp}^2 - \omega_{st}^2\big)^2}\right)v, \tag{102}$$

$$K = \left[1 + \left(\frac{\sinh(\omega_{st}T/2)}{\omega_{st}T/2} - 1\right)\right]^{-1}F_{ext}.$$

We look at the metabolic energy per step of the torso in the limit as $\omega_{st} \to 0$. In this limit, we find that the term in (101) in $C_2$ goes to zero. Taking the square of the integrand and noting that odd terms will integrate to zero, we find:

$$W_{st} \approx \varepsilon_{st}\big(\omega_{st}^2 - \omega_{sp}^2\big)^2 m^2 C_3^2 \int_{-T/2}^{T/2}\sinh^2(\omega_{st}t)dt$$
$$+ \varepsilon_{st}K^2\int_{-T/2}^{T/2}\cosh^2(\omega_{st}t)dt. \tag{103}$$

The integrals in (103) evaluate to:

$$\int_{-T/2}^{T/2}\sinh^2(\omega_{st}t)dt = \left(\frac{\sinh(\omega_{st}T)}{\omega_{st}T} - 1\right)(T/2),$$
$$\int_{-T/2}^{T/2}\cosh^2(\omega_{st}t)dt = \left(\frac{\sinh(\omega_{st}T)}{\omega_{st}T} + 1\right)(T/2). \tag{104}$$

In the limit we are looking at, we may make the approximation $sinh(\omega_{st}t) \approx \omega_{st}t + (\omega_{st}t)^3/6$. Therefore, in the limit as $\omega_{st} \to 0$, we find:

$$\int_{-T/2}^{T/2} \sinh^2(\omega_{st} t) dt \approx \omega_{st}^2 T^3 / 12,$$
$$\int_{-T/2}^{T/2} \cosh^2(\omega_{st} t) dt \approx T, \qquad (105)$$
$$C_3 \approx v / \omega_{st},$$
$$K \approx F_{ext}.$$

Thus, (101) approximates to:

$$W_{st} \approx \varepsilon_{st} \omega_{sp}^4 m^2 v^2 T^3 / 12 + \varepsilon_{st} F_{ext}^2 T. \qquad (106)$$

Evaluating $\omega_{sp}^4$ using the definition in (20) and noting that $s = vT$, we find:

$$W_{st} \approx \left(\varepsilon_{st} g^2 m^2 / 12 L^2\right) s^3 / v + \left(\varepsilon_{st}\right) F_{ext}^2 s / v. \qquad (107)$$

**Appendix 6**

The metabolic energy per step is associated with the motion of the swing foot is given in (71); it is:

$$W_{foot} \approx \varepsilon_{sw} \mu^2 \int_{-T/2}^{T/2} \left(6\kappa_2 t + C_3 \omega_{sw}^2 \sinh(\omega_{sw} t)\right)^2 dt. \qquad (108)$$

The constants are:

$$\kappa_2 = -\left(\frac{(\omega_{sw} T)^2}{3 T^3}\right) \sinh(\omega_{sw} T / 2) C_3,$$
$$C_3 = \left(\frac{12 T}{\left(12 + (\omega_{sw} T)^2\right) \sinh(\omega_{sw} T / 2) - 6(\omega_{sw} T) \cosh(\omega_{sw} T)}\right) v. \qquad (109)$$

We look at the metabolic energy per step of the torso in the limit as $\omega_{st} \to 0$. In this limit, we find that the term in (108) in $\kappa_2$ goes to zero. Taking the square of the integrand, we find:

$$W_{foot} \approx C_3^2 \varepsilon_{sw} \omega_{sw}^2 \mu^2 \int_{-T/2}^{T/2} \sinh^2(\omega_{sw} t) dt. \qquad (110)$$

The integral in (110) evaluates to:

$$\int_{-T/2}^{T/2} \sinh^2(\omega_{sw} t) dt = \left(\frac{\sinh(\omega_{sw} T)}{\omega_{sw} T} - 1\right)(T / 2). \qquad (111)$$

In the limit we are looking at, we may make the approximation $sinh(\omega_{st} t) \approx \omega_{st} t + (\omega_{st} t)^3/6$. Therefore, in the limit as $\omega_{st} \to 0$, we find:

$$\int_{-T/2}^{T/2} \sinh^2(\omega_{sw} t) dt \approx \omega_{sw}^2 T^3 / 12. \qquad (112)$$

We now approximate the constant $C_3$ in (109), in this case we make the $sinh(\omega_{st}t) \approx \omega_{st}t$ and $cosh(\omega_{st}t) \approx 1 + (\omega_{st}t)^2/2$; we find:

$$C_3 \approx -\left(\frac{48}{\omega_{sw}^3 T^2}\right)v. \tag{113}$$

Combining (110), (112), and (113), we find:

$$W_{foot} \approx 192\varepsilon_{sw}\mu^2 v^2 / T. \tag{114}$$

Noting that $s = vT$, we find:

$$W_{foot} \approx 192\varepsilon_{sw}\mu^2 v^3 / s. \tag{115}$$